\documentclass[10pt,aps,prb,twocolumn,superscriptaddress]{revtex4}

\usepackage{epsfig}
\usepackage{dcolumn}
\usepackage{bm}
\usepackage{amsmath}
\usepackage{amsfonts}
\usepackage{latexsym}
\usepackage{amssymb}
\usepackage{color}
\usepackage{hyperref}
\usepackage[normalem]{ulem}

\usepackage{tikz,xcolor,hyperref}
\definecolor{lime}{HTML}{A6CE39}
\DeclareRobustCommand{\orcidicon}{%
	\begin{tikzpicture}
	\draw[lime, fill=lime] (0,0)
	circle [radius=0.16]
	node[white] {{\fontfamily{qag}\selectfont \tiny ID}};
	\draw[white, fill=white] (-0.0625,0.095)
	circle [radius=0.007];
	\end{tikzpicture}
	\hspace{-2mm}
}

\foreach \x in {A, ..., Z}{%
	\expandafter\xdef\csname orcid\x\endcsname{\noexpand\href{https://orcid.org/\csname orcidauthor\x\endcsname}{\noexpand\orcidicon}}
}

\begin{document}

\title{Impact of correlations on topology in Kane-Mele model decorated with impurities} 

\author{Jan Skolimowski\orcidA}
\email{jskolimowski@magtop.ifpan.edu.pl}
\affiliation{International Research Centre MagTop, Institute of
   Physics, Polish Academy of Sciences,\\ Aleja Lotnik\'ow 32/46,
   PL-02668 Warsaw, Poland}

\author{Wojciech Brzezicki\orcidB}
\email{brzezicki@magtop.ifpan.edu.pl}
\affiliation{Institute of Theoretical Physics, Jagiellonian University, Prof. S. \L{}ojasiewicza 11, PL-30348 Krak\'ow, Poland}
\affiliation{International Research Centre MagTop, Institute of
   Physics, Polish Academy of Sciences,\\ Aleja Lotnik\'ow 32/46,
   PL-02668 Warsaw, Poland}

\author{Carmine Autieri\orcidC}
\email{autieri@magtop.ifpan.edu.pl}
\affiliation{International Research Centre MagTop, Institute of
   Physics, Polish Academy of Sciences,\\ Aleja Lotnik\'ow 32/46,
   PL-02668 Warsaw, Poland}

\date{\today}

\begin{abstract}

We propose an effective model for the study of the interplay between correlation and topology by decorating the Kane-Mele model with a set of localized interacting orbitals hybridized to just one sublattice, breaking the inversion symmetry. We show that in the time-reversal symmetric case, the interplay between interactions and hybridization extends the stability of the topological phase and 
depending on the driving mechanism very different behaviors are observed after the topological phase transition (TPT). We discuss the fate of the TPT in presence of weak ferromagnetic order, by introducing a weak local magnetic field at the localized orbitals, which splits the two band inversion points.
One of the platforms to apply this model to are ferrovalley compounds, which are characterized by two independent band inversion points.
Understanding this family of materials is crucial for the development of the {\it valleytronics}. An alternative to spintronics, which uses valley polarization as opposed to spin degrees of freedom as the building block, promises great opportunities for the development of information storage. 

\end{abstract}
\maketitle

\section{Introduction}

In the last two decades, the topological phase transitions (TPT) have attracted considerable attention\cite {doi:10.34133/2020/7832610,CONTINENTINO2017A1}. They can be driven either by applying external factors to the system or by tuning the interplay of intrinsic factors. In the first category are TPTs caused by pressure\cite{doi:10.1126/sciadv.aav9771,PhysRevB.100.195138}, temperature\cite{PhysRevB.106.L081103} and magnetic\cite{PhysRevMaterials.4.044202,Wysokinski23} or electric field\cite{Islam2021}. Latter category includes studies exploring the impact of correlations\cite{Rachel_2018,PhysRevLett.104.106408}, long-range order\cite{PhysRevB.98.045133,PhysRevB.106.054209} or lattice geometries \cite{PhysRevB.100.121107}. In case of correlations an additional complication arises from the fact that most of the topological classifications were obtained for non-interacting systems\cite{PhysRevLett.95.146802,PhysRevB.93.195413}. Proper extension of topological invariant to many-body systems is still under debate\cite{PhysRevX.2.031008,PhysRevLett.131.106601,PhysRevB.108.125115}.
Nonetheless, the interest in studying the interplay between electronic correlations and topology has been staidly growing for many years and is still an active field\cite{PhysRevLett.120.040504,Wysokinski23,Dzsaber2022,Pizarro2020,PhysRevB.99.075158,Wagner2023}. In particular, we should mention the field of the flat bands in two-dimensional and three-dimensional\cite{PhysRevX.11.031017} systems. More specifically, the study of the magic-angle twisted bilayer graphene was addressed as a topological heavy-fermion problem with a relevant role of the electronic correlations\cite{PhysRevLett.129.047601}.

A compound with a single band inversion usually becomes more trivial as a function of the Coulomb repulsion since the Coulomb repulsion promotes the opening of a trivial band gap. This was shown in model studies using both mean-field\cite{PhysRevB.91.161107,PhysRevB.103.035125} and many-body \cite{PhysRevLett.116.225305} approaches. The same happens in the density functional theory approach if we use an exchange-correlation functional that has more electronic correlations such as modified Becke-Johnson\cite{Hussain_2022}, HSE or strongly constrained and appropriately normed\cite{PhysRevB.108.075150}. 

Additional complexity in the topological phase diagram of a system can arise if it has two band inversions at inequivalent $k$-points. In the case of two band inversion at different $k$-points with different gap closures as a function of the Coulomb repulsion a two-stage correlation-driven topological transition was reported\cite{Hussain23correlation}.
A more complex topological phase diagram can be obtained under time-dependent light irradiation\cite{doi:10.1021/acs.nanolett.2c04651}.
More in general, if we have a complex Fermi surface the electronic correlations are able to induce topology\cite{Xu2020,Ma2022correlation}. Going beyond the single-particle picture, several topological Mott insulators have been proposed\cite{Rachel_2018, Wysokinski23}. 

One of the possible platforms to validate these results is provided by the family of two-dimensional (2D) materials called ferrovalley compounds\cite{Tong2016}. These compounds have 2 inequivalent band inversions at $K$ and $K^\prime$. Due to the presence of two inequivalent band inversions, these systems can host a more complex correlation-driven TPT\cite{Hussain23correlation,PhysRevB.105.195312,D3CP01368E,Guo_2022}. 
One path to obtain the ferrovalley compounds is through the decoration of a monolayer with ad-atoms, which is also a promising avenue to achieve new topological phases\cite{PhysRevB.106.125151} or enhancing certain properties of already existing phases.\cite{jiang2023monolayer} Decoration of 2D materials allows for engineering a system that has specific hybridizations and interactions to achieve different model Hamiltonians, such as the one examined in this paper.\\

In this work, we propose and study an effective model describing a hexagonal lattice, in which spin-orbit coupling (SOC) and inversion symmetry breaking through hybridization with strongly correlated orbitals coexist. By construction, it could play the role of a minimal model applicable to the compounds, which can host ferrovalley physics. There the interplay between these two factors is believed to be responsible for a wide range of phenomena such as the Quantum Anomalous Hall (QAH) phase.
The hexagonal lattice with SOC is modeled using the Kane-Mele model (KMM) and the inversion symmetry-breaking hybridization is modeled through coupling between one of the sublattices of the KMM and localized correlated orbitals. In order to extend our studies to a ferrovalley case, in the second part of this paper we will introduce a local magnetic field acting only on the localized orbitals. This model differs from the standard model of topological Kondo insulators\cite{PhysRevLett.104.106408}, where SOC is at the correlated impurity sites enforcing symmetry constraints on the hybridization term and giving rise to topology. It is also different from the standard periodic Anderson model on KMM\cite{PhysRevLett.111.016402}, where the impurity is always half-filled thus promoting the emergence of Kondo physics. In Section \ref{sec_model} we introduce the Hamiltonian and the Cluster Perturbation Theory method (CPT) used to solve it. In Section \ref{sec_no_mag_field} we will discuss the phase diagram of the model Hamiltonian without time-reversal breaking local magnetic field. Lastly, in Section \ref{sec_mag_field}, we will discuss the effect of the local magnetic field on the TPT in the KMM model and how it relates to the two-stage TPT reported in {\it ab initio} studies of ferrovalley materials\cite{Hussain23correlation}.

\section{Model}\label{sec_model}
The model Hamiltonian discussed in this work written in the second quantization form consists of three terms:
\begin{equation}\label{H}    \mathcal{H}=\mathcal{H}_{latt}+\mathcal{H}_{hyb}+\mathcal{H}_{imp},
\end{equation}
\begin{multline}
    \mathcal{H}_{latt}=-t\sum_{<i,j>,\sigma} c^\dagger_{i,\sigma}c_{j,\sigma}-\mu \sum_{i,\sigma}c^\dagger_{i,\sigma}c_{i,\sigma}+\\
 +i t^\prime \! \sum_{\ll i,j\gg,\sigma,\sigma^\prime}\nu_{i,j} c^\dagger_{i,\sigma}s^z_{\sigma,\sigma^\prime}c_{j\sigma^\prime}
\end{multline}

\begin{equation}
   \mathcal{H}_{hyb}=J\sum_{i\in A, \sigma}(d^\dagger_{i,\sigma}c_{i\sigma} + H.c.),
\end{equation}
\begin{multline}
    \mathcal{H}_{imp}=\sum_{i\in A} U d^\dagger_{i,\uparrow}d_{i,\uparrow}d^\dagger_{i,\downarrow}d_{i,\downarrow}+\\+\sum _{i\in A,\sigma}(\epsilon_d +\sigma M) d^\dagger_{i,\sigma}d_{i,\sigma}
    -\mu \sum_{i\in A,\sigma} d^\dagger_{i,\sigma}d_{i,\sigma},
\end{multline}
where $c_{i,\sigma},d_{j,\sigma^\prime}$ are the fermionic annihilation operators of (itinerant) electrons in the orbitals forming the lattice and localized orbitals, respectively,  at sites $i$ and $j$ and with spin projection $\sigma$ and $\sigma^\prime$. The $\mathcal{H}_{latt}$ describes the standard Kane-Mele model\cite{PhysRevLett.95.146802, PhysRevLett.95.226801}, with real hopping $t$ between the nearest neighbours in the honeycomb lattice and a complex hopping $ it^\prime$ between next-nearest neighbors. The summations $<i,j>$ and $\ll i,j \gg$ reflect the range of each hooping term. 
Since the honeycomb lattice is bipartite we introduce capital Latin letters $A,B$ to distinguish between the two sublattices.
The KMM is a version of the Haldane model in which the time-reversal symmetry is recovered through the introduction of a spin-dependent sign of the complex hopping, hence the Pauli matrix $z$ ($s^z_{\sigma,\sigma^\prime}$) in the last term of $\mathcal{H}_{latt}$. The parameter $\nu_{i,j}=\pm 1$ in this term distinguishes between a clockwise and anti-clockwise movement of particles within each sublattice. Such hopping was shown to describe a spin-orbit coupling \cite{PhysRevLett.95.226801}. The $\mathcal{H}_{hyb}$ combines all terms that connect the lattice with correlated localized orbitals. In the following we will consider the decoration of only a single sublattice with localized orbitals, thus summation is over $i\in A$. Parameter $J$ quantifies the strength of the hybridization between the electrons in the localized and lattice orbitals. Finally, $\mathcal{H}_{imp}$ is a set of local Hamiltonians, each having a Hubbard-type interaction term characterized by the strength $U$, local potential $\epsilon_d$ and Zeeman-like term coupling impurity electrons to an external magnetic field $M$. The latter is introduced to mimic either intrinsic or externally driven instability of the system to form magnetic moments. The parameter $\mu$ in $\mathcal{H}_{latt}$ and $\mathcal{H}_{imp}$ is the chemical potential. It will be tuned to study the 2/3 and 1/3 filled cases for which the TPT is allowed, cf. text below for a detailed discussion. In the following, when $U\neq 0$ then $\epsilon_d=-U/2$ will be assumed. To set the energy scale we use $t=1$.
\begin{figure}
\flushleft
\includegraphics[width=0.5\textwidth]{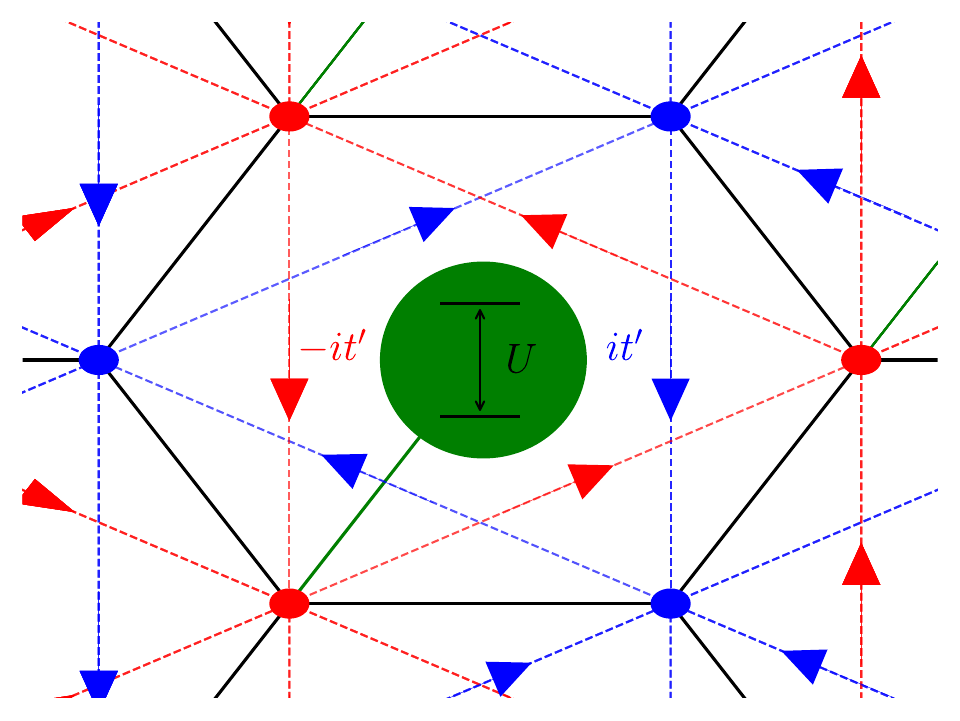}
\caption{Depiction of the lattice model described by Hamiltonian \ref{H}. The blue and red dots are the two non-equivalent lattice sites in the KMM, and the green dots are the localized orbitals. The black lines show standard hopping amplitude and blue and red dashed lines with arrows are the directional complex hoppings. Green lines are the hybridization $J$ terms between the lattice and the localized orbitals. At the latter, there is a Hubbard interaction illustrated as two levels separated by $U$. }
\label{lattice:fig}
\end{figure}

To solve the interacting Hamiltonian \ref{H} we will use the Cluster Perturbation Theory method. This method, introduced by Senechal et al. \cite{PhysRevLett.84.522}, builds upon the strong coupling expansion of the Hubbard model\cite{PhysRevLett.80.5389,Pairault2000} and allows to study both strong local correlation and kinetic effects\cite{PhysRevB.66.075129}. The $k$-dependent solution of the full Hamiltonian is built from an exact solution of an isolated cluster $\mathcal{G}_0(\omega)$ and matrix elements describing inter-cluster coupling in the reciprocal space $V(\vec{k}$) through RPA-like expression 
\begin{equation}\label{CPT}
    \mathcal{G}(\omega,\vec{k})=\mathcal{G}_0(\omega)\left[1-V(\vec{k})\mathcal{G}_0(\omega)\right]^{-1}
\end{equation}
The solution of the isolated cluster is usually obtained either analytically for simple subsystems or through the exact diagonalization method in the case of more complex clusters. The results presented in this work were obtained using the latter approach, with the isolated cluster being the unit cell of the lattice in Eq. (\ref{H}). It consists of three sites: two representing the non-equivalent sub-lattices of the KMM and the third site representing the localized orbitals, shown as blue, red and green circles in Fig. \ref{lattice:fig}. 
The localized nature of the interacting orbital in this model makes the inter-cluster coupling matrix $V(\vec{k})$ effectively a $2\times2$ matrix, as opposed to a $3\time 3$ matrix in a generic case.
The $2\times 2$ sub-matrix of $V(\vec{k})$ has almost the same elements as the $H(\vec{k})$ matrix for the bare KMM. The difference is in its off-diagonal elements, where the hopping between the atoms within the same unit cell is missing. The effects connected to this matrix element is included in $\mathcal{G}_0(\omega)$. The reduced dimensionality of $V(\vec{k})$ in the model studied here is responsible for the strong presence of local (intra-cluster) dynamics in the spectrum of the lattice systems.

The information about the topological state of the system will be read out from the {\it generalized} spin-Chern number ($C_s$) defined as a difference of {\it generalized} Chern numbers\cite{PhysRevX.2.031008} for each spin species. The need for $C_s$ comes from the time-reversal symmetry of this model at $M=0$, which leads to the cancellation of normal (generalized) Chern number similar to the standard quantum spin Hall system (QSH)\cite{PhysRevB.76.045302,PhysRevB.80.125327}. The $C_s$ invariant is an extension of the standard Chern number to many-body systems, through the utilization of the eigenstates of the effective Hamiltonian 
\begin{equation}\label{H_eff}
    \mathcal{H}_{\rm eff}(\vec{k})=-\mathcal{G}^{-1}(0,\vec{k}).
\end{equation}
It inherits the spin $S_z$ conservation symmetry after the 
full many-body Hamiltonian so it can
be written as
\begin{equation}\label{H_eff}
    \mathcal{H}_{\rm eff}(\vec{k})=\mathcal{H}_{\rm eff}^{\uparrow}(\vec{k})\oplus\mathcal{H}_{\rm eff}^{\downarrow}(\vec{k}).
\end{equation}
Now the generalized spin-Chern number $C_s$ can be defined as $C_s=C_s(\uparrow)-C_s(\downarrow)$ where $C_s(\sigma)$ is a standard Chern number of  $\mathcal{H}_{eff}^{\sigma}(\vec{k})$. It can be obtained by, e.g., Kubo formula:
\begin{equation}
C({\sigma})=\frac{1}{2\pi}\int_{BZ}d^2k\Omega_{\vec{k}}^{\sigma}
\end{equation}
with 
\begin{equation}
\Omega_{\vec{k}}^{\sigma}=\!\!\sum_{{n\le n_{F}\atop m>n_{F}}}\!\!\mathfrak{Im}\frac{\left\langle \psi_{\vec{k},\sigma}^{n}\right|\!\partial_{k_1}\mathcal{H}_{\rm eff}^{\sigma}\!\left|\psi_{\vec{k},\sigma}^{m}\right\rangle \!\left\langle \psi_{\vec{k},\sigma}^{m}\right|\!\partial_{k_2}\mathcal{H}_{\rm eff}^{\sigma}\!\left|\psi_{\vec{k},\sigma}^{n}\right\rangle }{\left(E_{\vec{k}}^{(n)}-E_{\vec{k}}^{(m)}\right)^{2}}.
\end{equation}
Here $|\psi_{\vec{k},\sigma}^{n}\rangle$ is an eigenstate of $\mathcal{H}_{\rm eff}^{\sigma}$ with energy $E_{\vec{k}}^{(n)}$ and $n_F$ is the number of occupied bands.
This is a non-zero quantized quantity in the topological phase of this model due to $S_z$ conservation \cite{PhysRevLett.97.036808, PhysRevB.80.125327}. In the text below, every time non-trivial topology is mentioned it will mean $C_s(\sigma) \neq 0$.
The accuracy of this approach in determining the topology on a many-body state is currently under debate \cite{PhysRevLett.131.106601,PhysRevB.108.125115}, yet it provides an estimate for the qualitative behavior of the topological phase diagram. In addition, it has been frequently used in similar studies of the interplay between correlations and topology\cite{PhysRevB.88.035113, PhysRevB.88.165132,PhysRevB.99.121113,PhysRevB.105.245127}. Using this approach will allow for qualitative comparison against these results.

We will consider only clusters that are geometrically the same size as the unit cell of the underlying KMM. Because the additional (localized) orbital does not have inter-cluster coupling, the effective Hamiltonian will inherit the same lattice symmetries as the KMM except for the inversion symmetry.

\section{Time-reversal symmetric case}\label{sec_no_mag_field}

We will start with the analysis of the Hamiltonian \ref{H} in the absence of magnetic field $M$ ($M$=0) i.e. the time-reversal symmetric case.
In the non-interacting limit, when $U=0$, the presence of localized orbitals can drive a topological phase transition through the interplay of two factors: (i) the hybridization strength $J$ and (ii) the position of the localized orbital level $\epsilon_d$. The latter sets the initial position of the flat band, of localized orbitals, relative to the dispersive bands formed by the itinerant electrons. The former mixes the two types of orbitals giving rise to $k$-dependence in the otherwise flat band at $\epsilon_d$. 

The three orbital unit cell of the model means, that the insulating phase can only be obtained for an even number of electrons. We will focus mainly on the $\frac{2}{3}$-filling and only mention the $\frac{1}{3}$-filling in cases, where it could lead to situations that are physically different. We will also consider the case of the flat band within the lattice electrons bandwidth, as later this configuration will allow us to explore the onset of Kondo screening once the correlations are included.

From the point of view of the TPT the $K/K^\prime$ points in the Brillouin zone (BZ), where the gap inversion in the KMM takes place, are crucial. The non-interacting Hamiltonian in the orbital basis for one of the spin channels is given by: 
\begin{equation}\label{non_int_H}
\mathcal{H}_{non,\sigma}(\vec{k})=\left[
    \begin{array}{ccc}
         \alpha_\sigma(\vec{k}) & 0  & 0\\
         0 & -\alpha_\sigma(\vec{k}) & J\\
         0 & J & \epsilon_d
         \end{array}
    \right].
\end{equation}

Depending on which $k=K,K^\prime$ point we look at the function $\alpha_\sigma(\vec{k})$ can have two values, $\alpha_\sigma(K)=\mathrm{sgn}(\sigma)3\sqrt{3}t^\prime $ or $\alpha_\sigma(K^\prime)=-\mathrm{sgn}(\sigma) 3\sqrt{3}t^\prime $.
At these two points the inversion symmetry-breaking hybridization $J$ introduces mixing between only two bands, while the third is effectively decoupled. As a result, the two hybridized bands repel each other, while the third band remains unchanged. Upon increasing $J$ one of the hybridized bands is eventually pushed across the gap to the unaltered band causing the change in the sign of the band gap and thus a TPT. The role of $\epsilon_d$ in this scenario is to set the initial energy spacing between the in-gap (localized) band and the conduction band. Hybridization alone is sufficient to drive the TPT, while $\epsilon_d$ is not. Worth stressing is, that in case the lattice electrons had an inverted gap at $J=0$, switching on the hybridization will bend the flat band at $K$ and $K^\prime$ points in opposite directions. As a consequence of the hybridization breaking the sub-lattice symmetry of a lattice with band inversion. On the other hand, if the lattice electrons were initially in a trivial state, i.e. due to additional sub-lattice splitting mass term, the band bending will induce curvature with the same sign at either of the two $k$-points. Thus, from the curvature of the initially localized band alone, one can deduce the topology of the underlying lattice. In addition, a hybridization that induces curvature with a different sign at $K$ and $K^\prime$ points in the flat band has a greater effect on its overall bandwidth, further differentiating these two cases. 

\begin{figure}
\flushleft
\includegraphics[width=0.49\textwidth]{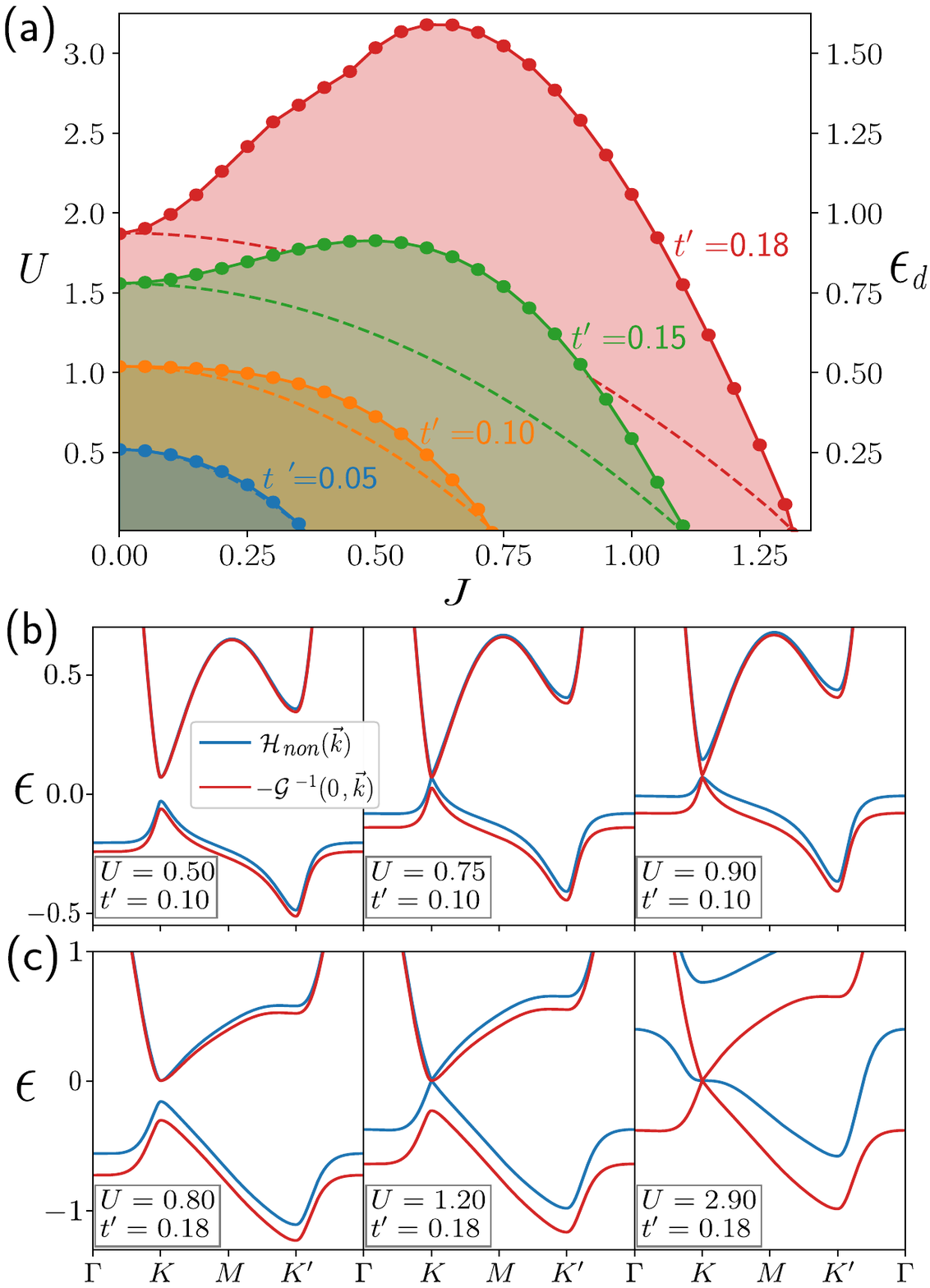}
\caption{(a) Phase diagram of the $\frac{2}{3}$ filled decorated KMM without a local magnetic field for various $t^\prime$ values. Filled areas indicate the stability regions of the topologically non-trivial phase with spin-Chern number $C_s=1$ for the corresponding NNN-hopping strength denoted in the same color. Dashed lines show the gap width of the corresponding un-correlated and un-decorated Kane-Mele model. (b) Low-energy band structure as a function of $U$ with $t^\prime$=0.10 and $J=0.4$ fixed for the non-interacting (solid blue line) and interacting case (solid red line). (c) Low-energy band structure as a function of $U$ with $t^\prime$=0.18 and $J=0.8$ fixed for the non-interacting (solid blue line) and interacting case (solid red line).
}
\label{phase_diag:fig}
\end{figure}

To illustrate and quantify the interplay between $J$ and $\epsilon_d$ in the non-interacting model we constructed the topological phase diagram, shown in Fig. \ref{phase_diag:fig}(a). The values of $\epsilon_d$ are indicated on the right-hand side of the figure and the coloured dashed lines show the TPT lines as a function of $J$ for various $t^\prime$. Each dashed corresponds to a certain value of $t^\prime$ displayed in this figure in the same color. The region below the dashed curves is the stability region of the topologically non-trivial phase, and above is the trivial insulator. At exactly $J=0$ there is no TPT since the localized orbitals are disconnected from the lattice. They are topologically trivial and changing their energy has no effect on the overall spin-Chern number of the system, which is made non-trivial only by the lattice electrons. Thus x-axis scale in Fig. \ref{phase_diag:fig}(a) starts from a very small but non-zero $J$. Outside this pathological case, as $J$ increases the stability region of the Chern insulator monotonically decreases irrespective of the $t^\prime$. This monotonic behavior reflects the cooperation of $\epsilon_d$ and $J$ in closing the topological gap.

Armed with a better understanding of the non-interacting limit we will now proceed to $U\neq 0$ cases, fixing the localized orbital energy level at $\epsilon_d=-U/2$. This choice of $\epsilon_d$ fixes the mid-point between the two levels of the localized orbitals to the center of the bare KMM. The resulting regions of stability of the topologically non-trivial phases for various $t^\prime$ are indicated in Fig. \ref{phase_diag:fig}(a) by shaded regions between the axes and continuous lines with circles. These lines define critical interaction strength $U_c(J)$ for which the gap is closed. Their colors, similarly to the non-interacting case, correspond to the colors of $t^\prime$ indicated in the figure. The $U$ values are indicated on the left-hand side of the figure. The scale on the left and right y-axis of Fig. \ref{phase_diag:fig}(a) was chosen in such a way as to allow comparison between the interacting model and a reference non-interacting model where the localized orbital contributes to the single energy level $\epsilon_d=U/2$. Their relation will be discussed in the next paragraph.

At small NNN hopping strengths ($t^\prime=0.05$) the dashed and solid lines match quite well. This means that one can draw a parallel between the interacting model with $\epsilon_d=-U/2$ and a non-interacting model with $\epsilon_d=U/2$. It is possible since at $J=0$, $U<6\sqrt{3}t^\prime$ and $\frac{2}{3}$ filling the spectrum of the system consist of a single band originating from the localized orbitals sandwiched between the bands of the lattice electrons. The disconnected localized orbitals contribute only one band due to the particular filling of the system, which makes the localized orbitals fully occupied. Through the analogy with a Hubbard atom, double occupancy results in a spectrum with a single level shifted from its bare value by $U$. The smallness of the gap ($\approx t^\prime$) makes a topologically non-trivial phase unstable already at relatively small values $J$, due to the limited energy window for the band bending. Weak hybridization does not allow for correlation effects to spill into the lattice and thus the system can still be effectively separated into an itinerant (lattice) and localized electrons. 

At larger $t^\prime$ the wider KMM gap allows for stronger hybridization $J$, which leads to the discrepancies between the dashed and solid lines in the intermediate region ($U\neq0$ and $J\neq 0$). In the limiting two cases of $J=0$ or $U=0$ the two lines have to match. At $U=0$ the two models are simply equivalent. At $J=0$ the (trivial) gap is closed when the upper Hubbard band crosses the KMM conduction band. Until then it remains below the Fermi level and the mapping to the effective model with a single flat band at $U/2$ is exact. But lack of hybridization makes this case uninteresting due to the lack of the TPT. Between the limiting cases (intermediate $J$), an increased stability of the topologically non-trivial phase is observed in the fully interacting model, when compared to the effective non-interacting model i.e. the shaded region extends beyond the dashed lines with the same color. For $t^\prime<0.15$ the shape of the stability region changes only quantitatively with $t^\prime$, keeping the same monotonic decrease with $J$. When $t^\prime > 0.1$ a qualitative change starts to appear. The topological phase first starts to become robust against larger $U$ as $J$ increases. Then the stability region develops a maximum at intermediate $J$ values, before dropping to the $U=0$ limit. As $t^\prime$ grows this maximum grows (moves to larger $U$) and shifts to larger $J$, indicating that stronger hybridization allows for screening of stronger interactions. But the maximal allowed $J$ value for the non-trivial phase is bounded by the $U=0$ limit, which is set by the $J=0$ lattice electrons gap width. Hence the observed behavior that the larger the $t^\prime$ the bigger the area between the solid and the corresponding dashed lines becomes. 

To illustrate this point further, in the two bottom panels of Figs. \ref{phase_diag:fig}(b,c) we compare the low energy spectrum of the effective Hamiltonian (Eq. \ref{H_eff}) and the corresponding non-interacting model. The former determines the topology of the interacting system \cite{PhysRevX.2.031008} and the latter plays the role of a reference case in which correlations are treated on a mean-field level. Fig. \ref{phase_diag:fig}(b) and Fig. \ref{phase_diag:fig}(c) show three regimes of $U$ for $t^\prime=0.1, J=0.4$ and $t^\prime=0.18, J=0.8$, respectively. These are weakly interacting (left column), critical $U$ for the effective non-interacting model (middle column) and critical $U$ for the fully interacting model (right column), respectively. The bottom bands (deep below the Fermi level) are not displayed in any of these cases as they are irrelevant from the point of view of the TPT. At weak interaction strengths $U=0.5, 0.8$, the displayed bands of the two Hamiltonians behave qualitatively the same. The conduction bands of $\mathcal{H}_{non}$ and $-\mathcal{G}^{-1}(0,\vec{k})$ follow each other closely around the $K$ point, and deviate slightly just on the $K\rightarrow K^\prime$ path, where hybridization effects come into play for a given spin channel. The biggest difference between the spectra is in the location of the valence band. Crucially, the direct gap at $K$ point in the spectrum of the non-interacting model is smaller than in the spectrum of $-\mathcal{G}^{-1}(0,\vec{k})$. In addition, comparing the left and the middle columns one can also see that the valence band is less prone to move towards the conduction band upon increasing $U$ once the correlation effects are accounted for. 
This points towards the reduction of the bare $U$ value by the hybridization term, through the delocalization of the charge in the interacting orbitals.

Comparison between the two spectra around the $K^\prime$ point, at first glance, would also suggest an effective increase of $J$, since the gap at this point in the BZ of $\mathcal{H}_{non}$ is smaller than of the $-\mathcal{G}^{-1}(0,\vec{k})$. However, the increase in $J$ around $K^\prime$ point is consistent with the reduced effect of $U$. The gap between the {\it two hybridized levels} ($\Delta_{non,\sigma}(\vec{k})$) of the Hamiltonian in Eq. (\ref{non_int_H}) is given by 
\begin{equation}
    \Delta_{non,\sigma}(\vec{k})=2\sqrt{J^2+\left(\frac{\alpha_\sigma(\vec{k})+\epsilon_d}{2}\right)^2}.
\end{equation}
Since $\alpha_{\sigma=\uparrow}(k=K^\prime)=-3\sqrt{3}t^\prime$ and $\epsilon_d=U/2$ the second term in the square root increases as U decreases. As a result the fully interacting model, with screened interactions has a larger gap in the spin-up channel at $K^\prime$ point. 
Contrasting the $U$ values displayed in the two rightmost columns in the two bottom panels, which shows the gap closing of $ \mathcal{H}_{non}$ and $-\mathcal{G}^{-1}(0,\vec{k})$, confirms that for larger $t^\prime$ this effect is even stronger extending the stability of the topological phase from $U=0.9$ at $t^\prime=0.1$ to $U=2.9$ at $t^\prime=0.18$.

\begin{figure}
\flushleft
\includegraphics[width=0.48\textwidth]{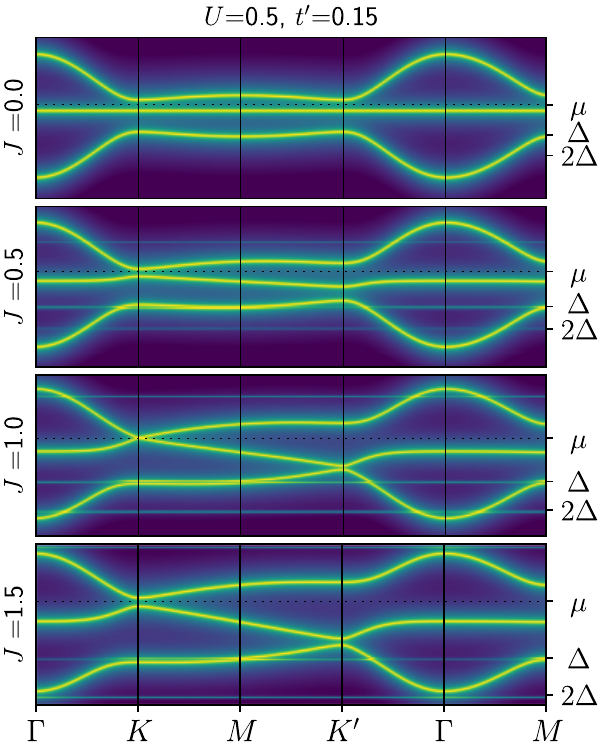}
\caption{Momentum resolved spectral function $A_\uparrow(\omega,k)$ along the high-symmetry points for $U=0.5$ and $t^\prime$=0.15. We tune the value of $J$ across the gap closing starting from the topological phase at $J$=0 up to $J=1.5$ where the system becomes trivial. On the y2-axis (righthand side), we indicate the Fermi level $\mu$ and the values of $\Delta=-\sqrt{J^2+t^2}$ and 2$\Delta$.}
\label{single_spin_band_TI:fig}
\end{figure}

The robustness of the conduction band due to the hybridization effect is not synonymous with the rise of the Kondo physics \cite{Hewson_book,doi:10.1080/000187300243345, PhysRevLett.111.016402}. The latter would lead to the formation of a narrow band pinned to the Fermi level \cite{PhysRevLett.111.016402}. In this model, the Kondo screening is, in most cases, suppressed due to near double occupancy of the interacting orbitals. For the values of $t^\prime$ examined in this work, the interaction strengths for which the topological phase is stable are below the underlying KMM bandwidth and the system is outside the so-called Kondo regime \cite{PhysRevB.62.5657}, in which the interacting orbitals are half-filled. As a result, the conduction band has a reasonably large dispersion and is not pinned to or crossed by the Fermi level. The non-trivial topology at 2/3 -filling differentiates this from a standard Kane-Mele-Kondo-lattice model \cite{PhysRevLett.111.016402} in which the hybridization term does not introduce inversion symmetry breaking. As a consequence the TPT takes place for electron densities in the interacting orbitals closer to half-filling, bringing the Kondo physics to the forefront.

We now proceed to the analysis of the spectral features of the full $\mathcal{G}(\omega,\vec{k})$ across the topological phase transition in two cases: (i) $J$-driven ($U\ll 1$)and (ii) $U$-driven ($J\ll 1$). These two have to be separated since the trivial states one could reach can be differentiated through analyzing the distribution of spin-Chern number between each filled band.

In Fig. \ref{single_spin_band_TI:fig} four $k$-resolved spectral functions for the spin-up channel along the high symmetry points of the BZ are displayed. These spectra are for a system with $t^\prime=0.15 $, weak interactions at the localized orbitals ($U=0.5$) and different hybridization strengths $J$ across the TPT. The corresponding spin-down channel results can be obtained through the time-reversal symmetry operator since in the absence of $M$ it is the symmetry of the system.
The top panel shows the spectrum without the hybridization ($J=0$). As anticipated it consists of two dispersive bands separated by a gap of $6\sqrt{3}t^\prime$ at $K$ and $K^\prime$ points, and a single flat band of the localized orbitals between them. The lack of a hybridization term allows for the interpretation of the dispersive bands as originating from the KMM. The middle band comes from the doubly occupied localized orbitals, underlined by the lack of dispersion. As mentioned above, it follows the behavior of a doubly occupied Hubbard model in the atomic limit. The double occupancy is the result of $\frac{2}{3}$ filling, hence the flat band is $U/2$ away from the KMM gap center. In the case of $\frac{1}{3}$ filling the localized orbitals would have been empty and thus also contribute a single band to the full spectrum. Its energy is lowered by $U$ with respect to the one displayed in the top panel of Fig. \ref{single_spin_band_TI:fig}.

After the introduction of a non-zero $J$ the spectrum of $\mathcal{G}(\omega,\vec{k})$ becomes more complex with features that can be divided into two sets. The first one consists of bands with orders of magnitude larger spectral weights, we will refer to them as main bands. These are the states directly connected to the eigenstates of the effective Hamiltonian from the Eq. (\ref{H_eff}). In the other set are excitations with much smaller spectral weights. We will refer to them as secondary bands. They are mostly flat and either isolated or have anti-crossings with the main bands. In the latter case, narrow bands are formed as a result. These excitations appear only when $U$ and $J$ are non-zero and their spectral weight is directly related to the hybridization strength.  The flatness of these states is a clear indication of their origin from intra-cluster dynamics.

The evolution with $J$ of the low energy part of the main bands reproduces the analysis made previously for the eigenstates of the $\mathcal{H}_{eff}(\vec{k})$. Upon increasing $J$, from zero, the initially flat conduction band becomes dispersive, due to mixing between mobile non-interacting lattice electrons and localized interacting electrons. 
Its curvature increases with $J$ and has the opposite sign at the $K$ and $K^\prime$ points, as a consequence of the presence of an {\it inverted gap} in the underlying KMM. Hence in Fig. \ref{single_spin_band_TI:fig} the band repulsion is between the flat valence band and the conduction band at one $k$-point and between the flat valence band and the main band below it at the time-reversal symmetric partner of that $k$-point.
In the opposite spin channel, the band bending is in exactly the opposite direction due to the time-reversal symmetry. The band, which does not participate in the anti-crossing, does not feel the change in the hybridization strength and remains at its $J=0$ position. At the critical strength of $J=J_c$ the repulsed and the unaffected bands cross and the TPT takes place (cf. third panel from the top in Fig. \ref{single_spin_band_TI:fig}).

The non-zero interaction strength plays also an important role in the shape of the stability region of a topological phase. At $J=0$ it controls how far away from the center of the inverted gap of the lattice the localized orbital level is occurring.

The lack of particle-hole symmetry in the topological phase has an interesting consequence. The off-center (within the KMM gap) placement of the interacting orbitals means, that as the hybridization strength increases and the band bending takes place the gap at $K/K^\prime$ points is closed for different $J$ values. For the set of parameters presented in Fig. \ref{single_spin_band_TI:fig}, it is first closed at $K$ point and later at $K^\prime$ point. After the first gap closing the system already becomes a trivial insulator and the second gap closing does not change the topology of the system anymore. 
But a closer inspection of the "generalized spin-Chern number" for each of the bands separately of the effective Hamiltonian in Eq. (\ref{H_eff}) reveals how the topological invariant jumps between them after each gap reopening. At small $J$ the middle band is trivial and the two outer bands carry an opposite topological invariant. After the first gap reopening the topological invariants of the two highest energy bands swap. Since the system is $\frac{2}{3}$ filled the total spin-Chern number of the occupied bands becomes zero. The second gap reopening, below the chemical potential, leads to the trivialization of each band separately. In the case of $\frac{1}{3}$ filling the spectra would be particle-hole and $\vec{k}\rightarrow-\vec{k}$ transformed, since the localized orbital would be empty. As a result, the gap for spin-up electrons will be first closed at $K^\prime$ and later for the $K$ point, but the overall behavior of the spin-Chern number would be qualitatively the same.
As mentioned earlier, the increase in $J$ strength results also in the appearance of flat band excitations, secondary bands, with relatively small spectral weight. They are located roughly at multiples of $\Delta=-\sqrt{J^2+t^2}$, which is the characteristic energy scale of the charge fluctuation within the unit cell cluster of the lattice in Fig. \ref{lattice:fig}. This further confirms their origin from local dynamics within the unit cell. Due to electron filling (2/3 or 1/3) not being at the particle-hole symmetry point of this system these flat bands are not visible in the same way in the particle spectrum ($\omega>\mu$) as they are in the hole part ($\omega<\mu$) of the spectrum. Best illustrated by the excitation at $\omega=\Delta$. This flat band crosses the broad dispersive band in the hole region of the energy spectrum around its upper edge. It has no counterpart in the electron part of the energy spectrum. This cannot be said about a pair of flat bands formed at higher energies, which are crossing the outer band edges of both main bands with large dispersion. 

\begin{figure}
\flushleft
\includegraphics[width=0.48\textwidth]{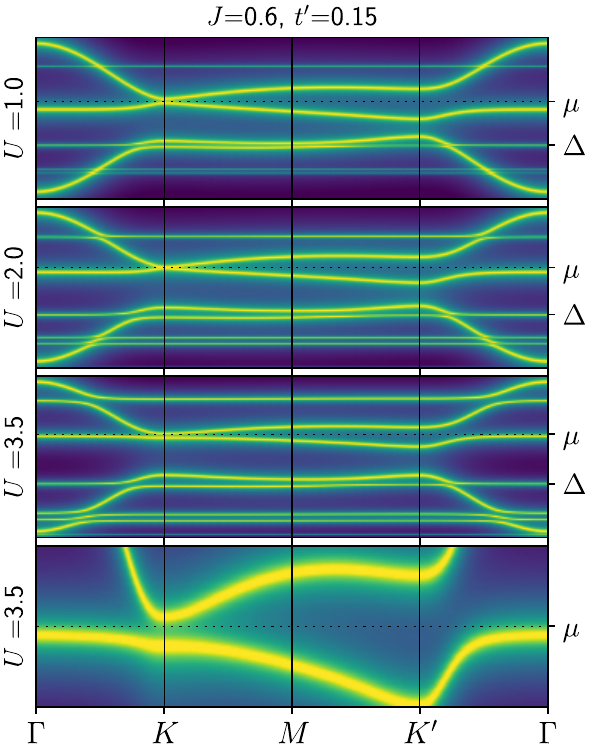}
\caption{Momentum resolved spectral function $A_\uparrow(\omega,k)$ along the high-symmetry points for $J=0.6$ and $t^\prime$=0.15. We tune the value of $U$ across the gap closing starting from the topological phase at $U$=1.0 up to $J$=3.5 where the system is a narrow gap trivial insulator. Bottom panel shows a zoom-in on the low energy part of spectra, highlighting the formation of a narrow gap around the $K$ point. On the y2-axis (right-hand side), we indicate the Fermi level $\mu$ and the values of $\Delta=-\sqrt{J^2+t^2}$.}
\label{single_spin_band_TI_variousU:fig}
\end{figure}

We turn now to the analysis of correlation-driven behaviors. Figure \ref{single_spin_band_TI_variousU:fig} illustrates the evolution of the spectral function across the TPT as the interaction strength $U$ is increased, while a non-zero (weak) hybridization is fixed at $J=0.6$. 
From the point of view of the effective non-interacting model, one would expect the gap between the conduction and valence bands to be closed through a shift of the latter across the gap. In contrast to the previous case, where it was caused by the valence band gaining dispersion.

The comparison between the three spectral functions displayed in Fig. \ref{single_spin_band_TI_variousU:fig} confirms that this different mechanism driving the gap closure is at play. Due to this fact, the gap is closed only at either $K$ or $K^\prime$ point depending on the spin species, but the time-reversal symmetry does not allow for spin-dependence in the critical $U$. The bandwidth of the valence band is not experiencing enhancement with increasing $U$, thus the gap below the Fermi level, at the TR symmetric $k$-point, is never closed. 

The comparison between the spectra in two top panels of Fig. \ref{single_spin_band_TI_variousU:fig} shows that initially the valence band undergoes a reduction in the bandwidth accompanied by only a very small decrease of the gap at the $K$ point. Eventually, upon increasing $U$ the topological gap closes and a trivial one opens. The re-opened gap, shown for $U=3.5$ in two bottom panels of Fig. \ref{single_spin_band_TI_variousU:fig}, is narrow and indirect. Further increase of $U$ only flattens the conduction band. This evolution with $U$ is not expected from the effective non-interacting model, cf. blue curve in the rightmost plot in the bottom row. There the conduction band undergoes an increase in the bandwidth after the TPT, as it contains parts of the flat band at $\Gamma$ point and the valence band of the KMM at $K$ point. The observed behavior in the interacting model can only be explained as an onset of the Kondo regime. For the top panel weak interactions ($U< 6\sqrt{3}t^\prime$) is limiting the hybridization effect to renormalization of the gap. At $U=2$ the interacting orbital levels move deep into the KMM bands, and its occupancy drifts away from double occupancy and towards the half-filling. That is when the simple single-particle picture starts to slowly break down\cite{PhysRevB.62.5657}. The rise of Kondo physics leads to the creation of a many-body resonant state pinned to the chemical potential\cite{Hewson_book}. This state, in the case of a lattice of localized interacting orbitals, becomes split by the hybridization gap \cite{PhysRevB.61.12799}. Thus after the TPT the spectrum of the system has a narrow gap and an almost flat valence band. The width of the gap at the chemical potential is set by the Kondo coherence scale.

On top of the changes around the Fermi level flat bands form at higher energies, as in the previously discussed case. These secondary bands are also clustering around multiples of $\Delta$, reflecting their inter-cluster dynamics origin. The interaction strength, surpassing the hybridization, enhances their anti-crossing with the dispersive main bands and reveals the internal structure of the high-energy flat bands. The internal structure of the flat bands is an aftermath of the charge fluctuations at the interacting orbitals caused by $J$, doping it away from double occupancy, combined with their two-level spectra due to a charge gap. The spectral weight of each level is connected to the density of holes and electrons in the interacting orbitals, thus the internal structure reveals itself in the hole part of the spectrum. These effects were probably also present in the $U=0.5$ case, shown in Fig. \ref{single_spin_band_TI:fig}, but the smallness of $U$ and the broadening factor used in calculating $\mathcal{G}(\omega,\vec{k})$ did not allow to resolve them. From the comparison between the results for $U<J$  in Fig. \ref{single_spin_band_TI:fig} and $U>J$ shown in Fig. \ref{single_spin_band_TI_variousU:fig}, one can see that the spectral weight in the high energy flat bands remain small in the former while increases in the latter. These bands are the precursors of the charge excitation peaks of the interacting orbitals characteristic in the spectrum of the periodic Anderson model \cite{PhysRevB.61.12799,PhysRevB.62.5657}. As $U$ becomes the dominating energy scale, the spacing between them approaches $U$ and their interpretation becomes clear.

A TPT in an opposite direction, from a band insulator to a topological insulator, can be also observed in systems upon adding a sufficiently strong Semenoff-type sub-lattice splitting mass term \cite{PhysRevLett.53.2449,PhysRevLett.61.2015}. In this geometry, the lack of an {\it inverted gap} makes the action of increasing $J$ or $U$ similar. In either case, the valence band is pushed towards the Fermi level, leading to an eventual gap closing and reopening. There still is some increase in the bandwidth of the valence band due to increased hybridization, but because the band repulsion is between the two fully occupied bands at any point of the BZ this effect much weaker. The absence of a large increase in the bandwidth of the valence band caused by $J$ means that a hybridization-driven second TPT, back to a trivial insulator, is not allowed. That would be the case if the $J$ term would allow for gap inversion between the two occupied bands with different Chern numbers. 

\section{ Local magnetic field}\label{sec_mag_field}

Lastly, we will analyze the impact of the local magnetic field acting on the electrons in the localized orbitals in the scenarios described above. This {\it orbital selective} field is introduced to mimic a possible ferromagnetic instability in the system. It can be either an intrinsic instability due to the system's dynamics, as reported for a similar model of decorated graphene \cite{Seki_2015}, or driven by an external field generated by the lack of inversion symmetry as in Janus systems \cite{Zheng23}. 

In this section we will focus on the $U$-driven transition since this case could show a similar transition to the reported in Ref. \onlinecite{Hussain23correlation}, from trivial to non-trivial and back to the trivial phase upon increasing interaction strength. In addition, the stability of the non-trivial phase around the transition point for the paramagnetic case relied on the emergence of the local Kondo physics. The introduction of a spin-symmetry breaking term, in theory, should strongly influence the stability of the low-energy physics\cite{PhysRevLett.85.1504}. From that point of view, the $J$-driven transition is less interesting since the interacting orbitals levels are within the bare KMM gap and the system overall is less susceptible to display many-body behavior.

\begin{figure}
\flushleft
\includegraphics[width=0.48\textwidth]{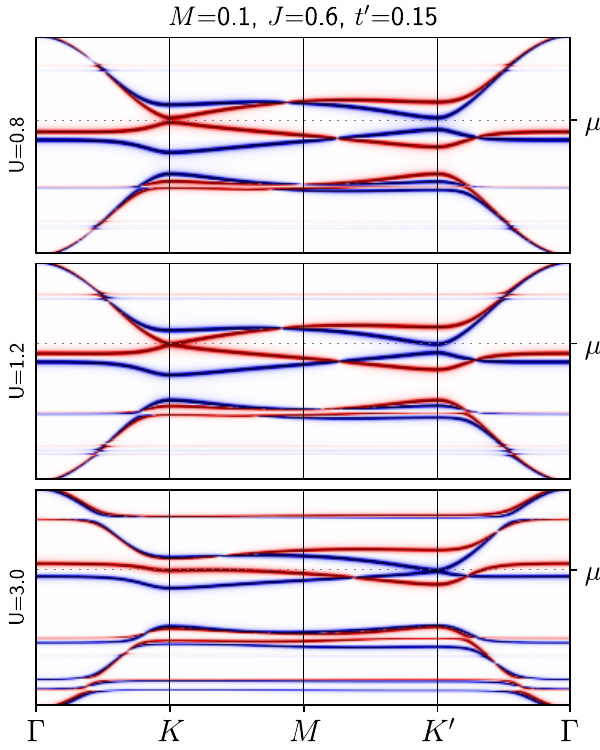}
\caption{Difference in the momentum resolved spectral function $A_\uparrow(\omega,k)-A_\downarrow(\omega,k)$ along the high-symmetry points for $M$=0.1, $J=0.6$ and $t^\prime$=0.15. Blue color indicates spin-down majority bands and red color indicates the spin-up majority bands. Interaction strength $U$ is varied across the gap closing at $K$ and $K^\prime$ points. (Top panel) Starting at $U=0.8$  the system is topological with in-equivalent band inversion points. (Middle panel) At $U=1.2$ the gap at $K$ point is closed and the system enters a pseudogap phase. (Bottom panel) At $U=3$ bands in the spin-down channel cross at the Fermi level $\mu$ (indicated on the right-hand side), which is inside the (narrow) band of spin-up excitations.}
\label{mag_field_spin_comp}
\end{figure}

Figure \ref{mag_field_spin_comp} shows the evolution of spectral functions for similar parameters as used in Fig. \ref{single_spin_band_TI_variousU:fig}, but in the presence of a small local magnetic field $M=0.1$. Since the time-reversal symmetry no longer holds Fig. \ref{mag_field_spin_comp} displays the difference in the spectral weights between the two spin channels along high symmetry points of the BZ, to show results for both spin channels in a concise way.

Blue color indicates the spin-down majority and red color the spin-up majority excitations. The lack of time-reversal symmetry is signaled by the different gap widths around the Fermi level at $K$ and $K^\prime$ points for the two spin species. Compared with the results for similar $U$ values without a magnetic field before the TPT, top panels of Fig. \ref{single_spin_band_TI_variousU:fig} and Fig. \ref{mag_field_spin_comp}, the gap around $K$-point is reduced for spin-up and increased at $K^\prime$ for spin-down electrons due to the Zeeman term. The magnetic field orientation is such that the valence band for spin-up electrons is shifted upwards and for spin-down electrons downwards. As a result, upon increasing $U$ first the gap at $K$-point in the spin-up channel closes and later the gap closer takes place at $K^\prime$ in the spin-down channel. 
Nonetheless, the system transitions only from a TI to a metallic state. The trivial gap is never opened, as the gap in one spin channel opens within the opposite spin channel band. 
Similar behavior was reported in Ref. \onlinecite{Hussain23correlation}, where the reopening of a gap around one $k$-point was initially overshadowed by the metallic behavior at its time-reversal symmetric partner in the BZ. In this model system, it is caused by the bottom of the spin-down conduction band at $K^\prime$ crossing the Fermi level at the same time the spin-up channel undergoes a TPT, as shown in the middle panel of Fig. \ref{mag_field_spin_comp}. This is not coincidental, but a consequence of the symmetries of the lattice and the sub-lattice symmetry-breaking hybridization. To understand that it is best to look at the spectrum of the effective non-interacting model, which was showing qualitatively similar behaviours to the full model for $U<6\sqrt{3}t^\prime$. In the previous analysis, we showed that around $K$ and $K^\prime$ points, there exists a level that is decoupled from the other two, thus unaffected by $U$ nor $J$. It is also not affected by $M$, which only acts directly at the interacting orbital and indirectly at the lattice orbital coupled to it. These decoupled levels before the TPT are the bottoms of the conduction band at $K$ and $K^\prime$ points for the spin-up and spin-down channels, respectively. As shown in the two top panels of Fig. \ref{mag_field_spin_comp} they are at the same energy. Because they do not feel the magnetic field this analysis holds true irrespective of the strength and sign of $M$. After a gap closing in each of the spin channels these fixed points jump to the valence band. The existence of these constraints on the energy levels around the $K/K^\prime$ points means that after the gap closing in the spin-up channel, the total spectral function cannot have a true gap. Since the bottom of the spin-down conduction band at $K^\prime$ point has to have the same energy as the top of the spin-up valence band at $K$ point. In a special case of an almost flat valence band, a pseudo gap is possible. The gap closing at $K^\prime$ is shown in the bottom panel of Fig. \ref{mag_field_spin_comp}. At this stage, the band touching at $K^\prime$ is fully within the spin-up metallic band. From this panel, one can also see that the Kondo physics is no longer present in the system. The gap at $K$-point in the spin-up channel is much larger than for the same parameters and larger $U$ in the paramagnetic case, cf. bottom panel in Fig. \ref{single_spin_band_TI_variousU:fig}. The discrepancy cannot be explained simply by the Zeeman field shifting the interacting orbital level. For the parameters displayed in Fig. \ref{single_spin_band_TI_variousU:fig} and Fig. \ref{mag_field_spin_comp}, that would account for increasing $U$ by $M/2$ which is still smaller than $U=3.5$ shown in the former figure. This means that the flat bands seen in the presence of the magnetic field originate from local moments forming within each spin channel and recovering the effective single-particle picture.

\begin{figure}
\flushleft
\includegraphics[width=0.48\textwidth]{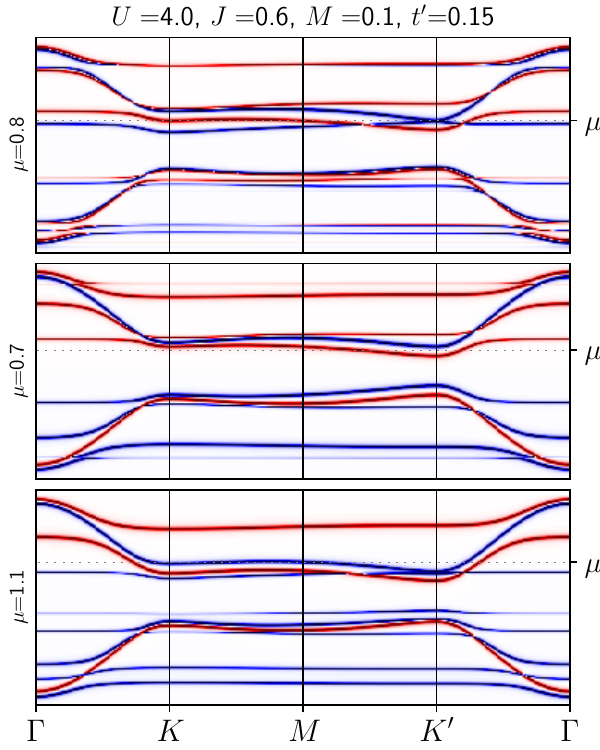}
\caption{Difference in the momentum resolved spectral function $A_\uparrow(\omega,k)-A_\downarrow(\omega,k)$ along the high-symmetry points for $U$=4.0, $M$=0.1, $J=0.6$ and $t^\prime$=0.15. The result for $\mu=0.8$ (top panel), $\mu=0.7$ (middle panel) and $\mu=1.1$ (bottom panel) are shown. Blue bands are the spin-down-majority, while the red bands are the spin-up-majority excitations. The position of the Fermi level $\mu$ is shown on the right-hand side.}
\label{mag_field_mu_change}
\end{figure}

To test this we analyze the influence of varying the chemical potential on the spectral function. If the Kondo physics would still be at play, the low energy excitations should adjust to the position of the chemical potential. In case the flat bands are simply from hybridizing the local moments with the lattice, changing $\mu$ should eventually drive spin-polarization and the vanishing of some flat bands in the spectrum. As shown in Fig. \ref{mag_field_mu_change}, for $U=4$ and the same parameters as in Fig. \ref{mag_field_spin_comp}, adjusting the chemical potential in the narrow window of $\delta \mu=0.3$ results in a drastic change of the band structure. This supports the {\it non-Kondo} origin of low-energy flat bands.

The top panel shows the spectrum of the system in case of the chemical potential pinned to the band crossing point at $K^\prime$ point, as a reference point. At low energies, the band structure in both spin channels behaves differently due to the trivialization of the gap in the spin-up channel (red color). However, the high-energy features show the same qualitative behavior, irrespective of spin direction, with only slight splitting due to the presence of $M$. Introducing a relatively small shift in the chemical potential, in any direction, leads to the breakdown of this resemblance at high energies due to changes in the orbital spin composition of the system. 
As shown in the two bottom panels of Fig. \ref{mag_field_mu_change}, reducing $\mu$ from $\mu=0.8\rightarrow \mu=0.7$ (middle panel) as well as increasing it from $\mu=0.8 \rightarrow \mu=1.1$ (bottom panel) results in a formation of the continuous spin-polarized band with dispersion from the decoupled KMM. The lack of localized levels crossing these bands is signaling the blocking of the intra-cluster charge fluctuation which was formed through mixing with one of the two local moment levels in the interacting orbital, as a result of its (partial) spin polarization. 
The spin-up band is located in the hole-part of the spectrum while the spin-down band is in the electron part of the spectrum, reflecting the direction of $M$. 

At $\mu=0.7$ the relative depletion of the spin-down electrons, compared to $\mu=0.8$, leads to disappearance of the the low energy excitations below the Fermi level in that spin channel. The system enters spin-polarized interacting orbital limit as the spectra in the spin-up and spin-down channels become connected by a reflection symmetry with regard to the gap center of the bare KMM. Above the KMM gap in the spin-down spectrum, only one continuous band survives, which is the upper bands of the KMM. Below the gap avoided crossings are clearly visible, which form due to the hybridization of the local moment level with the lattice.
For $\mu=1.1$ the chemical potential is moved into the upper band of the underlying KMM introducing more lattice electrons into the system, thus the two spin-resolved spectra become very distinct. 
In the spin-up channel it has only three bands while in the spin-down channel, the band structure is more complex. The reduction of the number of bands to only three and the absence of any flat bands means that the local charge fluctuations are frozen and the system can be directly mapped onto an effective non-interacting model with the localized orbitals level at $\epsilon_d=U/2$. The spin-down sector, on the other hand, has multiple avoided crossings in the occupied part of the spectrum but the electron part of the spectrum remains similar to spin-polarized case. This indicates that the inter-cluster dynamics is not yet trivialized.

\section{Conclusions}

We have studied the interplay between topology and electronic correlations in the decorated Kane-Mele model where we have two in-equivalent band inversions at $K$ and $K^\prime$ points. We demonstrate that in the presence of the time-reversal symmetry the interplay between the possible band inversion at two $k$-points, the electronic correlation and the sub-lattice symmetry breaking hybridization extends the topological region of the phase diagram beyond $U$ values predicted by a simple mean-field treatment. Despite the transition being at 2/3 (1/3) filling the Kondo physics arises in the system only at large $U$, making the topologically trivial phase a narrow gap insulator. 

By introducing an artificial magnetic field, mimicking ferromagnetic ordering at localized orbitals, we showed that a TPT at different $U$ values for each spin channel is not allowed in this model due reminiscence of time-reversal symmetry around $K$ and $K^\prime$ points forcing the position of one of the original KMM bands for the two spin species to be at the same energy. In addition, such a magnetic field would immediately destroy the Kondo physics making the system susceptible to small variation in chemical potential. Our results reveal that a non-trivial topology can survive in a system after decorating with correlated orbitals. Ferromagnetic ordering of these orbitals could block the opening of a trivial gap. 

We expect that the inclusion of hybridization to the other sublattice could allow for observation of separate TPT at the two non-equivalent band inversion points. Exploration of this topic combined with similar studies using methods that more accurately account for embedding local charge fluctuation in the lattice \cite{PhysRevLett.91.206402, PhysRevLett.101.186403, PhysRevB.107.235150}, could help better understand the properties of this model and explore its relevance to the ferrovalley compounds.

\vspace{5mm}

\begin{acknowledgments}
We thank M. Wysokiński and C. Mejuto Zaera for the useful discussions. The work is supported by the Foundation for Polish Science through the International Research Agendas Programme co-financed by the European Union within the Smart Growth Operational Programme (Grant No. MAB/2017/1). 
J. S. is supported by the National Science Centre (Poland) OPUS 2021/41/B/ST3/04475. 
J.S. and W.B. acknowledge support by Narodowe Centrum Nauki (NCN, National Science Centre, Poland) Project No. 2019/34/E/ST3/00404. 
We acknowledge the access to the computing facilities of the Interdisciplinary Center of Modeling at the University of Warsaw, Grant g91-1418, g91-1419 and g91-1426 for the availability of high-performance computing resources and support. We acknowledge the CINECA award under the ISCRA initiative  IsC99 "SILENTS”, IsC105 "SILENTSG", IsB26 "SHINY" and IsB27 "SLAM" grants for the availability of high-performance computing resources and support. We acknowledge the access to the computing facilities of the Poznan Supercomputing and Networking Center Grant No. 609.
\end{acknowledgments}

\bibliographystyle{apsrev4-2}
\bibliography{biblio}

\end{document}